\documentclass[nofootinbib,twocolumn,aps]{revtex4-1}
\usepackage{longtable,dcolumn,epsfig,color}
\usepackage[switch*,modulo]{lineno}

\begin{document}
\title{Paschen's law studies in cold gases}
\author{R. Massarczyk}
\affiliation{Los Alamos National Laboratory, Los Alamos, NM, USA}
\author{P. Chu}
\affiliation{Los Alamos National Laboratory, Los Alamos, NM, USA}
\author{C. Dugger}
\affiliation{Los Alamos National Laboratory, Los Alamos, NM, USA}
\affiliation{New Mexico Institute of Mining and Technology, Socorro, NM, USA}
\author{S.\,R. Elliott}
\affiliation{Los Alamos National Laboratory, Los Alamos, NM, USA}
\author{K. Rielage}
\affiliation{Los Alamos National Laboratory, Los Alamos, NM, USA}
\author{W. Xu}
\affiliation{Los Alamos National Laboratory, Los Alamos, NM, USA}
\affiliation{Department of Physics, University of South Dakota, Vermillion, SD, USA}

\begin{abstract}
The break-through voltage behavior over small gaps has been investigated for differing gap distances, gas pressures, and
gas temperatures in nitrogen, neon, argon and xenon gases. A
deviation from Paschen's law at micro gap distances has been found. The
break-through behavior of the fill gas in colder environments was also tested. A significant shift of the curve relative
to the results at room temperature was observed. This behavior can be explained by combining Paschen's law and the
ideal gas law.
\end{abstract}


\maketitle
\section{Introduction}
Paschen's law~\cite{Paschen1889} describes the electric discharge between two conductive materials as a function of gap distance ($d$) and the pressure ($p$) of
the intervening gas. For large gaps ($d > $ 100$\,\mu$m), the ``Paschen curve''
describes the break-through voltage $V_B$ as,
\begin{equation}
 V_B = \frac{B\,pd}{ln(A\,pd)-ln[ln(1+\frac{1}{\gamma_{SE}})]}.
\label{Eq_1}
\end{equation}
In addition to the two fit parameters A and B, the Second Townsend coefficient ($\gamma_{SE}$), which describes the mean
number of generated secondary electrons per ion, is also needed. Since this coefficient is often poorly known, the
parameters $A$ and $\gamma_{SE}$ are usually combined into a parameter $A'$, and Eq.\,\ref{Eq_1} becomes,
\begin{equation}
 V_B = \frac{B\,pd}{ln(A'\,pd)}.
\label{Eq_2}
\end{equation}

Equation\,\ref{Eq_1}  predicts very high break-through voltages for very small products of $pd$ as verified by the
breakdown measurements in vacuum or at very low pressures where no sparking is observed. As $pd$ is increased, a minimum
in $V_B$ is found and as $pd$ is further increased,  $V_B$ also increases above this minimum value. Some studies
\cite{Boyle1955, Ono2000, Peschot2014, Go2010} have found a deviation from Eq.\,\ref{Eq_1}  because
field emission at the cathode reduces the break-through voltage for very small gaps.

Paschen's law impacts a wide range of instruments in various environmental conditions and several studies have been
performed during recent years.
For example, possible breakthrough, often referred to as sparking or
voltage breakdown, can negatively influence rover missions on Mars \cite{Manning2010}, the performance of
accelerators \cite{Blaz2012}, the performance of semi-conducting detectors, along with other systems. The primary
motivation for the measurements in this study is the impact on the performance of detectors in a cold gas environment.

High-purity semi-conducting germanium detectors (HPGe) are widely used in nuclear and particle research, nuclear safety,
and security. Traditionally, these detectors are operated in a high-vacuum environment at low temperatures.
Low-radioactivity-background research,
such as neutrinoless double-beta decay, has inspired new detector designs and
technologies. In two recent experiments, the {\sc Majorana demonstrator}\cite{Abgrall2014} and {\sc
Gerda}\cite{Agostini2013}
have assembled arrays of germanium detectors built to search for this rare decay. The two collaborations use different
approaches to operate detectors. {\sc Majorana} uses a vacuum cryostat and {\sc Gerda} uses a cryostat filled with
liquid
Argon. (It has been previously shown that it is possible to
operate germanium detectors in liquid Nitrogen \cite{Klapdor2006} and liquid Argon \cite{Heider2009}.) Each method has
its advantages and disadvantages, but in both, the goal is to cool and operate detectors in a very
clean and extremely low-background environment. A large-scale experiment is being proposed for the future, and various
R$\&$D efforts are underway to improve detector performance.

One suggested concept is to operate the germanium detectors in a gaseous environment which could simultaneously
eliminate nearby neutron-producing high-Z materials and liquid Ar with its long-lived radioactivities. It may also be
possible to exploit scintillation light from a gas as a background diagnostic. Since HPGe detector operate at a
significant potential difference (typically up to 5\,kV), the conductor separation within the gas is a design
consideration. Furthermore, a decrease of $V_B$ with higher
temperature ($T$) for fixed gap distances has been observed in Nitrogen and air \cite{Uhm2003}. Other work has
considered the $V_B$
behavior in cold gases. Reference \cite{Fujita1978} claims no shift of $V_B$ over $d$ in a closed cold system, with
constant volume and therefore, constant particle
number. These findings, however, are valid only for specific setup. The HPGe detector design may
require an open, floating system with gases at different temperatures, pressures and possibly phases. Reference
\cite{Gerhold1979} came to similar conclusions, but
noted that generalized Paschen curves are required to describe $V_B$ as a function of
$T$. The work claims that the curve should not depend on $pd$ but the areal density, cf. Fig 8
in Ref. \cite{Gerhold1979}. This areal density can be interpreted to influence  the mean free path of
electrons in the gap gas. Since few published reports are available,
dedicated measurements are required for our gas candidates and electrode design constraints.

The Paschen curve has been measured for several popular fill gases, e.g. Nitrogen~\cite{Hackam1969a} and
Neon~\cite{Miller1964, Hackam1969, Bhattacharya1976}. In this work we measured $V_B$ in Nitrogen, Argon, Neon, and
Xenon.  All 4 gases produce scintillation light when irradiated, although in very different quantities. The main focus
in this study lies on measurements at
higher pressures and distances. Distances of a few mm are usually used to isolate parts under high-voltage in detector
design. Higher pressures are desirable due to the corresponding higher densities and
therefore in higher scintillation light production. On the other hand the pressure should not be so high as to produce a
liquid phase.

In the next section, this work gives an overview of the apparatus used for the Paschen curve studies. The results for
the
measurements with different fill gases at room temperature and in a cold environment are presented in Section
\ref{Sec_results} and the results are discussed in the last section of this manuscript.

\section{Paschen Curve Test Stand}
A steel cryostat was attached to a gas-handling system that maintained the pressure of various gases. The cryostat was
located within a dewar and permited cooling the cryostat with liquid
Nitrogen. Different gas temperatures could be achieved by varying the cryogen level.
Neon purified to 99.999$\%$, Nitrogen purified to 99.9$\%$, and Xenon and Argon both purified to 99.99$\%$, were
used in this study.
The gas system was able to handle gases in a pressure range from 1300 Torr down to 10$^{-5}$ Torr. The digital pressure
gauge was calibrated for Nitrogen
with a precision of less than a percent down to the mTorr range. However, an analogue gauge was used to monitor the
pressure for all gases. This gauge works properly for all gases mentioned with a precision of about 10$\%$ down
to about 100 Torr.\\

\begin{figure}[t!]
 \includegraphics[width=0.9\columnwidth,keepaspectratio=true]{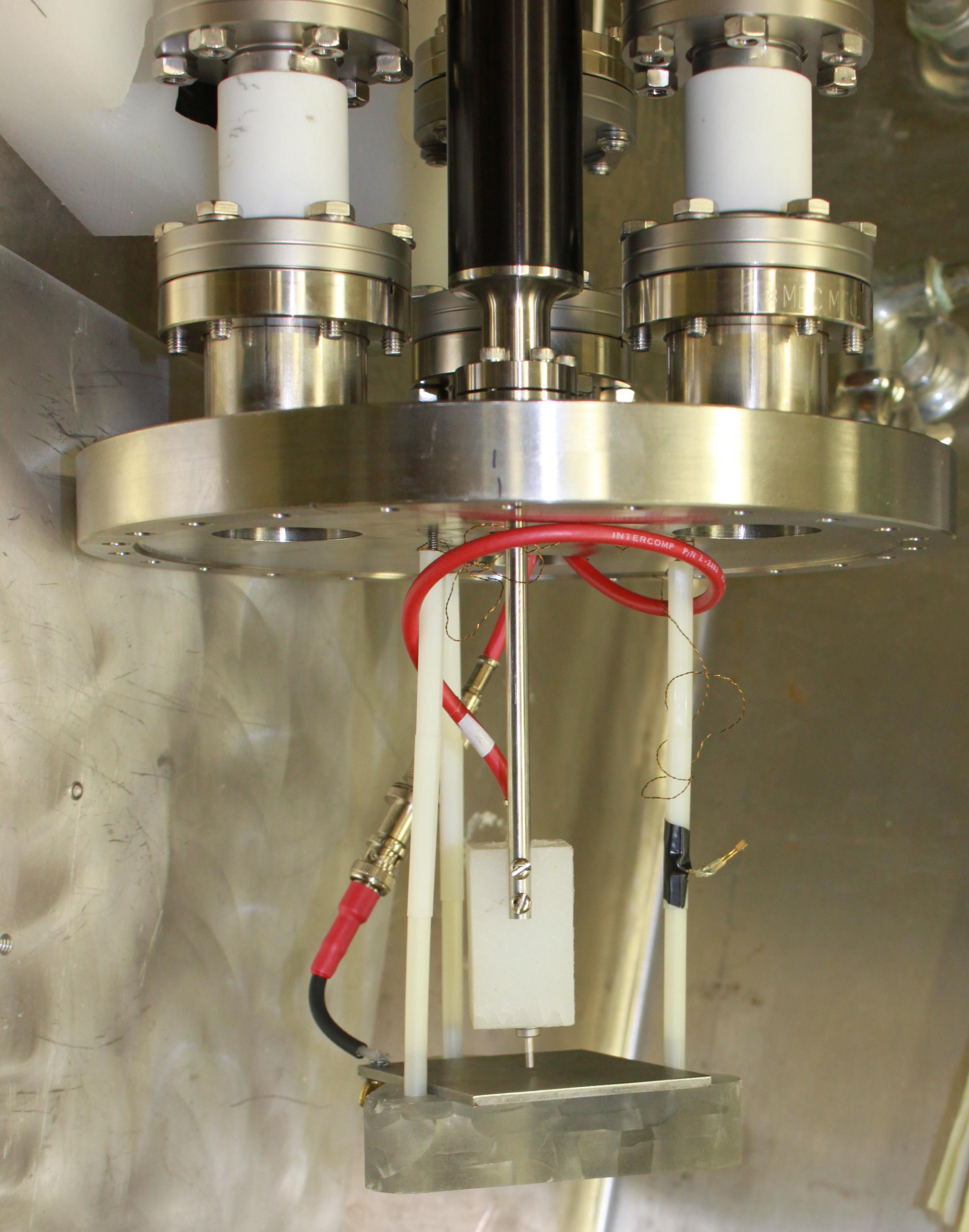}
 \caption{(color online) Photograph of the inner setup including support structure and cables which
are attached to the flange of the cryostat. The red cable supplies high voltage. White plastic parts are used for
isolation. }
 \label{Fig1}
\end{figure}

\begin{figure}[t!]
 \includegraphics[width=1.0\columnwidth,keepaspectratio=true]{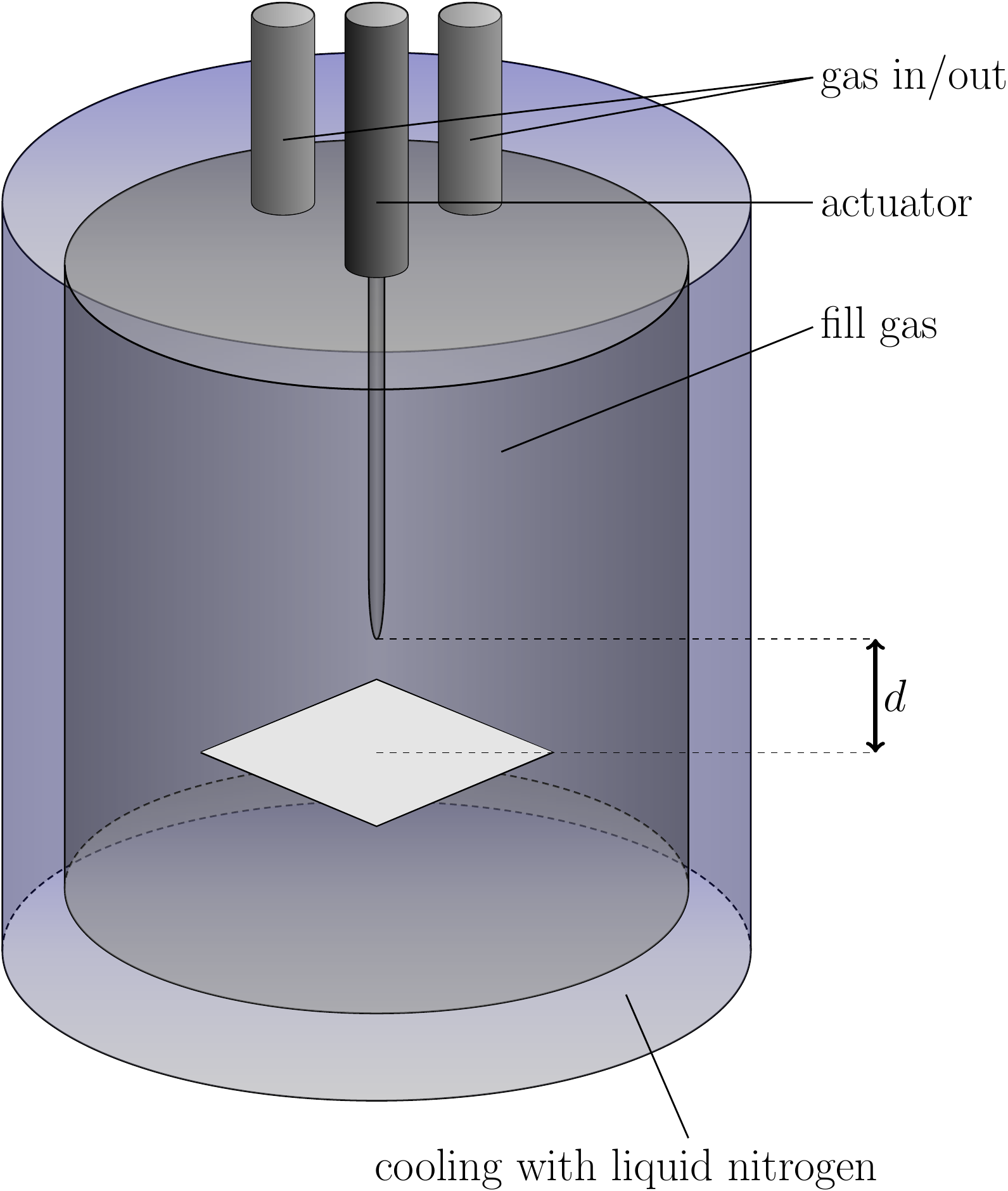}
 \caption{(color online)Schematic overview on the inside of the test stand. The gap width $d$ represents the shortest
break
through distance between the pin and plate.}
 \label{Fig2}
\end{figure}

An overview on the inside of the cryostat, is shown in Figs.\,\ref{Fig1} and\,\ref{Fig2}. Inside the cryostat, a
5x5\,cm$^{2}$ stainless steel plate was installed. The plate was held by a polyethylene structure and connected to a
feedthrough via a grounded
high-voltage (HV) cable. A pin\footnote{Ceramtec — $\#$8630-02-W} was located over
the plate in a fixed in position within a
Polytetrafluoroethylene (PTFE) block that was attached to an actuator\footnote{MDC, Linear motion feed-through}. The
linear motion of the actuator allows precision gap measurements between 0 and 10\,cm in 0.0005-cm steps.
The distance between the plate and pin mounts, and the surrounding cryostat is larger than 2.5\,cm. Therefore, 2.5\,cm
represents an upper limit to the gap distance. However, this limit was never reached by the available high
voltages.

\begin{figure}[t!]
 \centering
 \includegraphics[width=0.95\columnwidth,keepaspectratio=true]{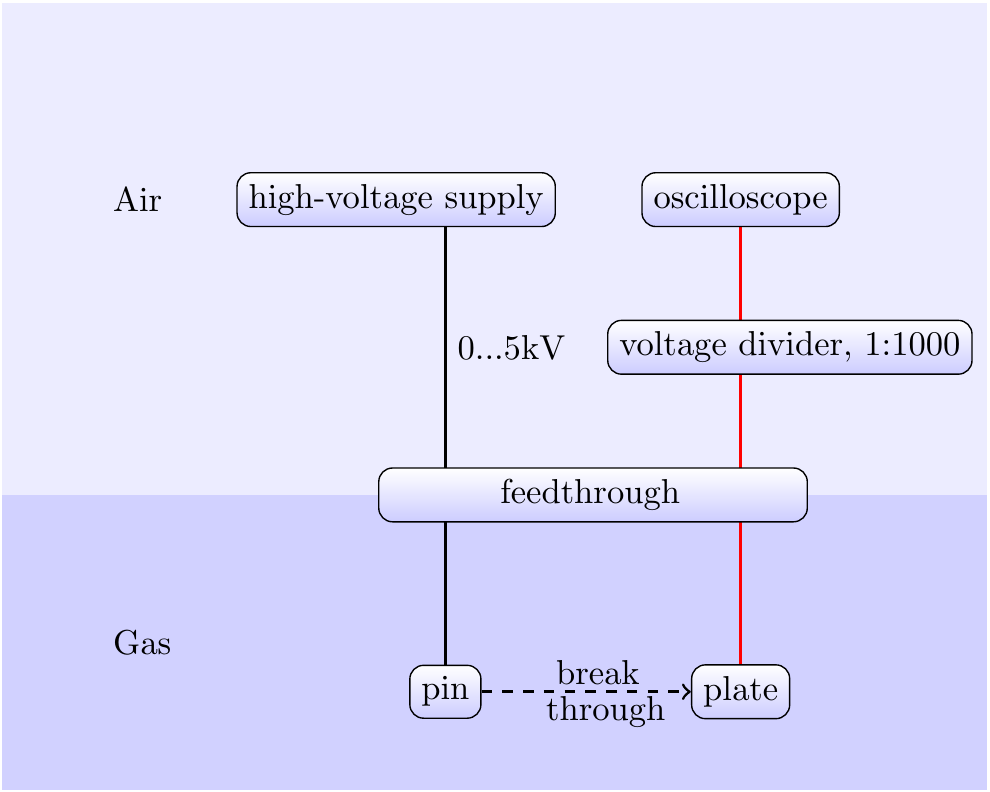}
 \caption{(color online) Wiring diagram of the setup used for the break-through measurements.}
 \label{Fig3}
\end{figure}

An overview of the wiring diagram is given in Fig.\,\ref{Fig3}. A 5-kV HV power supply\footnote{Bertran Model 325}
is used to provide voltages between 0 and 5\,kV with a precision of 0.01\,kV. The pin is connected over a
cable to the feedthrough flange. The plate is wired to the feedthrough and then to a voltage
divider. The voltage divider uses
 a 1.05-M$\mathrm{\Omega}$ and 1.02-k$\mathrm{\Omega}$ resistor to obtain a voltage decrease factor of 1000.
 The resulting maximum voltage of 5 V permitted the use an oscilloscope\footnote{LeCroy WaveRunner 610Zi} to observe the
$V_B$ signal. It was possible to change the cabling so that the high-voltage was set to the plate with the pin at
 ground. Several data points have been double checked in this inverse setting to verify that the breakthrough is
occuring between the pin and plate.
\\

The temperature inside the cryostat was measured by temperature sensors and read out by a temperature
monitoring system.\footnote{LakeShore Temperature Monitor}.

A series of measurements were performed. Before each measurement we waited until the system was stable and in
thermal equilibrium. After each cool-down the system was warmed and stayed at room temperature for a minimum of one day.
A vacuum pump was actively pumping the system to remove all residuals from the previous fill gas during the warm up and
between fills. The following measurements were made in this study.

\begin{itemize}
 \item \textbf{Nitrogen, room temperature (295K)}. This measurement was used as a  control due to previous measurements.
 The whole system was pumped out
and filled with Nitrogen. The break-through voltage was measured for different gaps at standard pressure (750 Torr).
After that the
system was evacuated, the gap was adjusted again to zero, and a new
measurement with different gaps was performed. For each gap setting, the $V_B$ was measured 5 times and
the values presented represent the mean. A fit of this data determined the Paschen-curve parameters in Eq.\,\ref{Eq_1}.

 \item \textbf{Nitrogen, cooled (150K)}. The system was filled with Nitrogen and cooled down to 150K. During the cool
down, Nitrogen gas was refilled inside so that the pressure remained normal. After
a conditioning time of about 3 to 4 hours during which no change in the temperature
was observed, the measurements were performed following a similar procedure as above. During these
measurements the cryostat was surrounded by liquid Nitrogen but not submerged.

\item \textbf{Nitrogen, cooled (80K)} Here, the whole cryostat was submerged in liquid Nitrogen. This cooled
 the gas inside the cryostat to the boiling temperature of Nitrogen.

 \item \textbf{Neon, Argon, Xenon, room temperature (295K), respectively}. After the series of measurements with
Nitrogen, the fill gas was changed. As described above, the system was evacuated before being filled with the next
gas. Various gap distances were measured at different pressures. Due to the limited range of the pressure gauge, we
were not able to measure gaps at very low pressure as previously done in Nitrogen.

 \item \textbf{Neon, Argon, Xenon, cooled (150K), respectively} Using the different fill gases we obtained the
same temperature inside the cryostat as before (with the Nitrogen fill gas) by surrounding the cryostat with liquid
Nitrogen but not submerging it.
\end{itemize}

While the cryostat was filled with Argon we proceeded to cover it with liquid Nitrogen to achieve 80K. After the
cool down we observed an instability in the deduced $V_B$ value at a fixed pressure and distance. We
expect that we created a solid phase next to a gas phase while cooling down at standard pressure (750 Torr). This
additional phase creates local pressure fluctuations and causes an unstable readout. As a consequence, we
limited the measurements of Xenon, Argon and Neon to temperatures of 150 K. To achieve these the cryostat has not to be
submerged all the way over the cover with liquid nitrogen and the gas purity is ensured.

\section{Results}
\label{Sec_results}

\subsection{Nitrogen}
\begin{figure}[t!]
 \centering
 \includegraphics[width=1.0\columnwidth,keepaspectratio=true]{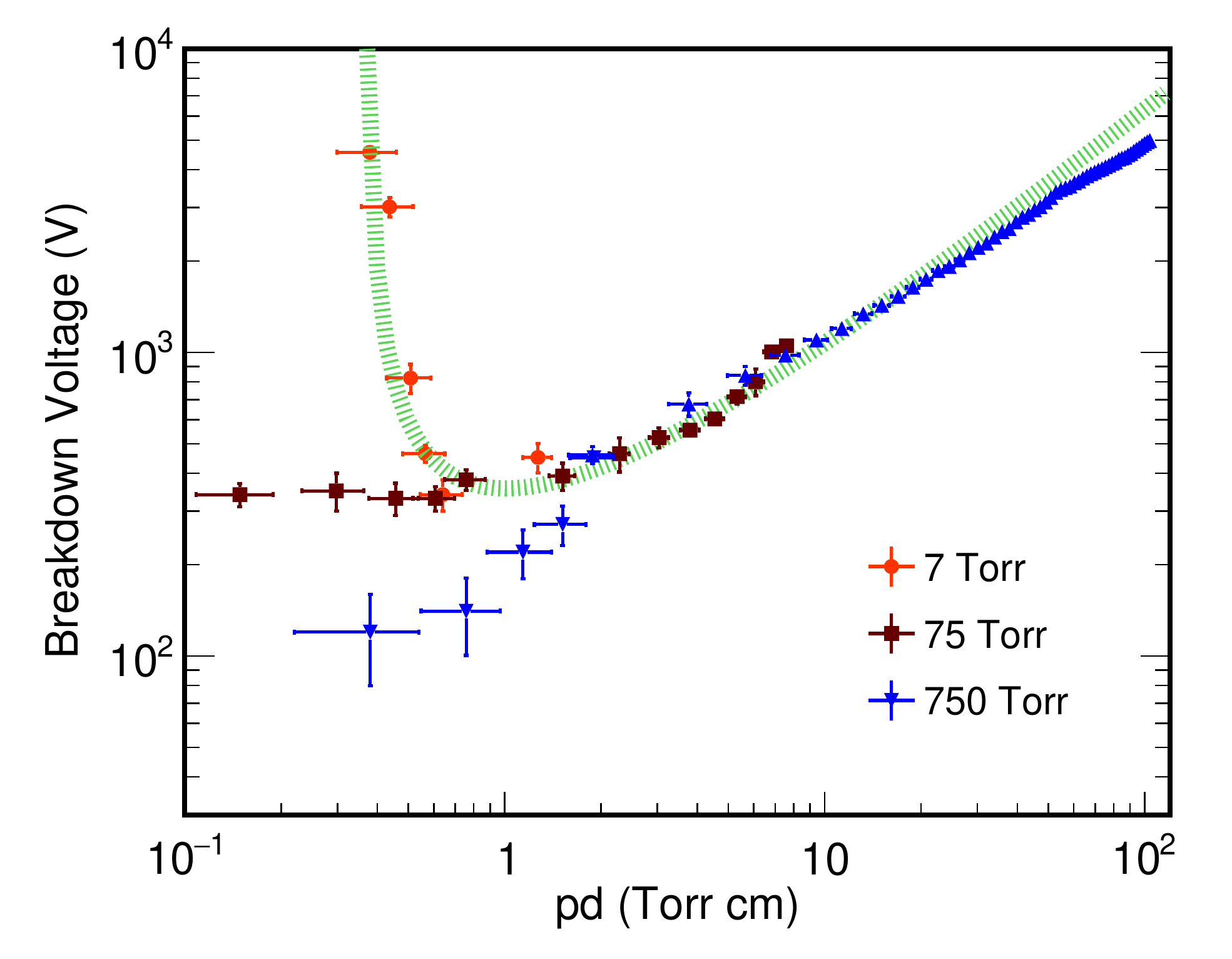}
  \caption{(color online) Results for the break-through voltage measurements in Nitrogen for room temperature (upper
figure) and cooled (lower figure). Different colors indicate different pressure
settings (750 Torr (blue) , 75 Torr (brown), 7 Torr (orange)). The green dashed line represents the curve
with the parameter given in Table\,\ref{Table1}. }
 \label{Fig4}
\end{figure}

We measured $V_B$ in Nitrogen for different pressures and temperatures. The results for room
temperature are shown in Fig.\,\ref{Fig4}. In addition to the statistical uncertainty of individual measurements, each
data point had an systematic uncertainty in voltage on the order of 0.01\,kV which reflects the precision of the HV
supply readout.

For standard pressure we saw no $V_B$ increase at
lower $d$ as predicted by Paschen's law. For $d$ smaller than 25 $\mu$m we found a continuous decrease in $V_B$, in
agreement
with the results presented by Refs. \cite{Boyle1955,
Ono2000, Peschot2014, Go2010}. The reason for this decrease are micro-discharges which may also initiate breakthrough at
small distances for low field strengths. This region was investigated by using lower pressures. On the one hand side,
lower pressures allow the same $pd$ value with larger gap sizes. On the other hand side, an investigation for the same
small gaps was done with lower pressure to determine the presence of small $V_B$ values in another pressure regime.
At low pressure (7\,Torr) larger gaps result in larger $V_B$ for the same value
of $pd$. In addition, for this $pd$ and $T$, a clear
rise in $V_B$ for low gap, as indicated by Paschen's law, was seen. We fit that data to Eq.\,\ref{Eq_1}. The results are
given in Table.\,\ref{Table1} in which a comparison of different literature values is also given. In order to improve
the
results of the fit we
restricted the fit to the parameter $A'$ and $B$, cf. Eq.\,\ref{Eq_2}. Therefore, no extra fit of the Townsend
parameter
is needed. Our fit results agree well with previous measurements.

At intermediate pressures (75 Torr) a small increase in $V_B$ for $pd<$0.5 Torr\,cm can be seen. For the smallest gap
measured at this pressure the microdischarge effect seems to play a role again and no further increase was observed.
Similar behavior has been found for other gases \cite{Matejik2015}. \\

After cooling, a new series of measurements was performed at $T=$150\,K. A clear shift of $V_B$
towards lower $pd$ was found, cf. Fig.\,\ref{Fig5}. That is at lower $T$s, $V_B$ is higher than for higher $T$s for
the same $pd$.
As also seen at room temperature, we did not observe an upturn at lower values of $pd$, as
$V_B$ continued to decrease with smaller $d$ at standard pressure. The violation of Paschen's law at small $pd$ was also
observed in a cold environment.
After cooling the system down to 80 K we again observed a further shift of $V_B$ towards lower $pd$
values.

\subsection{Neon, Xenon and Argon}
At standard room temperature and pressure the three gases seem to have the same general behavior as Nitrogen. While
measuring at small $d$, no increase of $V_B$ was found. This agrees with our findings
in Nitrogen and is well documented for Argon in Ref.\,\cite{Matejik2015}. At lower $p$, our measurements at the smallest
$d$
seems to indicate an upturn for Neon as shown in the data set in Fig.\,\ref{Fig5}. The data of Argon and Xenon
show a change in the slope at lower $pd$ which might also indicate an upturn in this region.
For the data sets at room temperature, fits of
Eq.\,\ref{Eq_2} were done. The results are given in Table\,\ref{Table1}. Again, reasonable agreement with previous
work was found.
The uncertainties of the fits are larger since no upturn influences the fit parameters.

After cooling, data for all three gases were taken for different pressure settings, except Xenon which is not entirely
gaseous at these temperatures and pressures. As mentioned before, the gas pressure
was kept constant at around 750 Torr during the cool down. The phase diagram of Xenon indicates that a liquid or even a
solid phase was produced. At 150 K, however, Xenon is in the gas phase near 250 Torr, where we conducted the
measurements.

\begin{table*}[t]
  \begin{tabular}{l l | c c c c c c c c}
    & & this work
	& Ref.\,\cite{Berzak2006}
	& Ref.\,\cite{unknown}\footnotemark[1]
    & Ref.\,\cite{Bazelyan1998}\footnotemark[1]
	& Ref.\,\cite{Naidu1995}\footnotemark[2]
	& Ref.\,\cite{Carazzetti2009}\footnotemark[3]
	& Ref.\,\cite{Raizer1997}\footnotemark[3]\\
 \hline
 \hline
 & & & & & & & & &\\
 Air & $A'$ (cm Torr)    & -- & 2.9(2)  & 3.8 & 4.3 & 4.8 & -- & -- \\
     & $B$ (V /\,cm Torr) & -- & 380(20) & 649 & 365 & 576 & -- & -- \\
 & & & & & & & & &\\
 Helium & $A'$ (cm Torr)    & -- & 0.77(2) & 0.7 & -- & 0.68 & 0.69 & 0.60 \\
        & $B$ (V /\,cm Torr) & -- & 35(2)   & 60  & -- & 39   & 39   & 34 \\
 & & & & & & & & &\\
 \hline
 & & & & & & & & &\\
 Nitrogen & $A'$ (cm Torr)    & 2.94(18) & -- & 2.5 & -- & 4.0 & 2.97 & 3.69\\
          & $B$ (V /\,cm Torr) & 349(50)  & -- & 550 & -- & 374 & 274  & 341\\
 & & & & & & & & &\\
 Neon & $A'$ (cm Torr)    & 1.31(26) & 1.84(6) & -- & -- & -- & -- & -- \\
      & $B$ (V /\,cm Torr) & 92(32)   & 130(20) & -- & -- & -- & -- & -- \\
 & & & & & & & & &\\
 Xenon & $A'$ (cm Torr)    & 1.92(26) & 3.71(8) & -- & -- & -- & -- & -- \\
       & $B$ (V /\,cm Torr) & 224(38)  & 193(8)  & -- & -- & -- & -- & -- \\
 & & & & & & & & &\\
 Argon & $A'$ (cm Torr) & 3.15(30)   & -- & 3.1 & -- & 3.02 & 2.64 & 3.57\\
       & $B$ (V /\,cm Torr) & 154(31) & -- & 320 & -- & 152  & 133  & 180 \\
 & & & & & & & & &\\
 \hline
 \hline
\end{tabular}
\footnotetext[1]{ $A$ is given, $A'$ is calculated by using $\gamma_{SE}$ = 0.001.}
\footnotetext[2]{ $pd_{min}$ and $V_{min}$ are given. $A'$ and $B$ are calculated using Eq.\,\ref{Eq_2} and its
derivation.}
\footnotetext[3]{ $B$ is given, $A'$ is calculated using Eq.\,\ref{Eq_2} and $pd_{min}$ and $V_{min}$ from
Ref.\,\cite{Naidu1995}}

\caption {Results for the fit with Eq.\,\ref{Eq_2} on the room-temperature data of this work and parameters form other
references. All
fits were performed for the data set at room temperature. The parameters for air and Helium are given for comparison to
illustrate the fluctuations in literature. Uncertainties for the literature values are given, if available.
Uncertainties of our measurements are given by the uncertainties of the fit.}

\label{Table1}
\end {table*}

\section{Discussion}
\label{Sec_discussion}
\begin{figure}[t!]
 \includegraphics[width=0.9\columnwidth,,keepaspectratio=true]{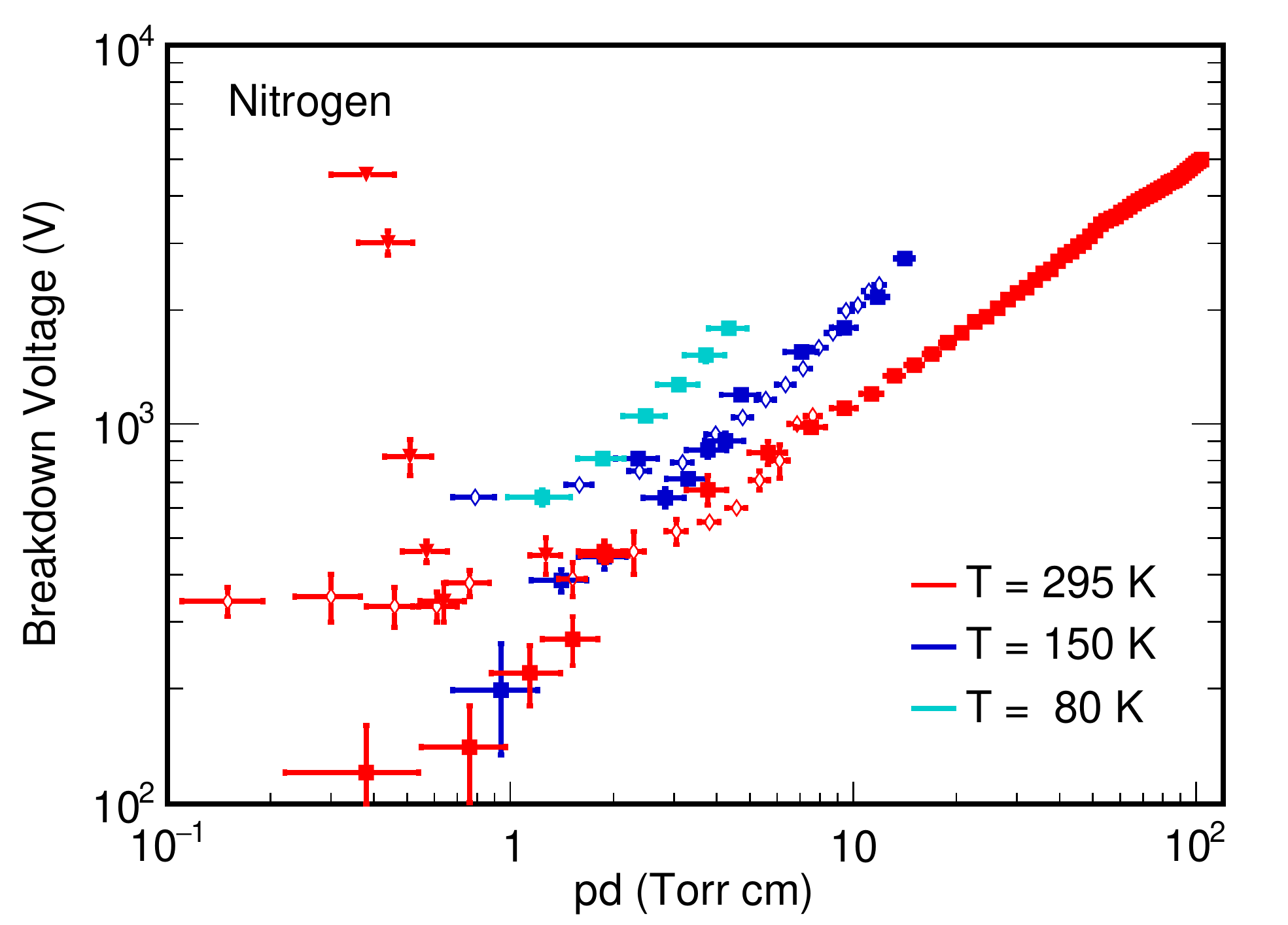}
 \includegraphics[width=0.9\columnwidth,,keepaspectratio=true]{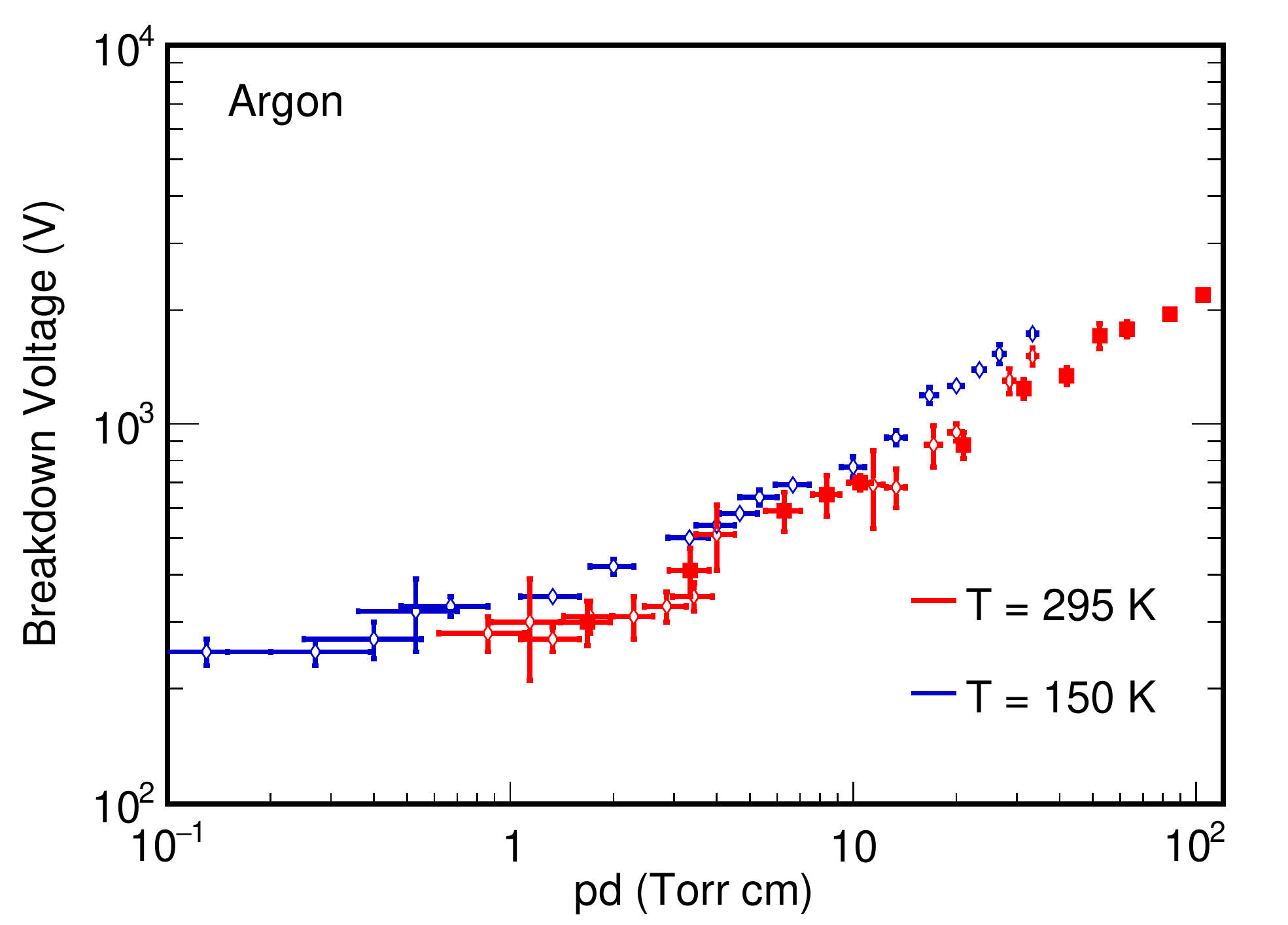}
 \includegraphics[width=0.9\columnwidth,,keepaspectratio=true]{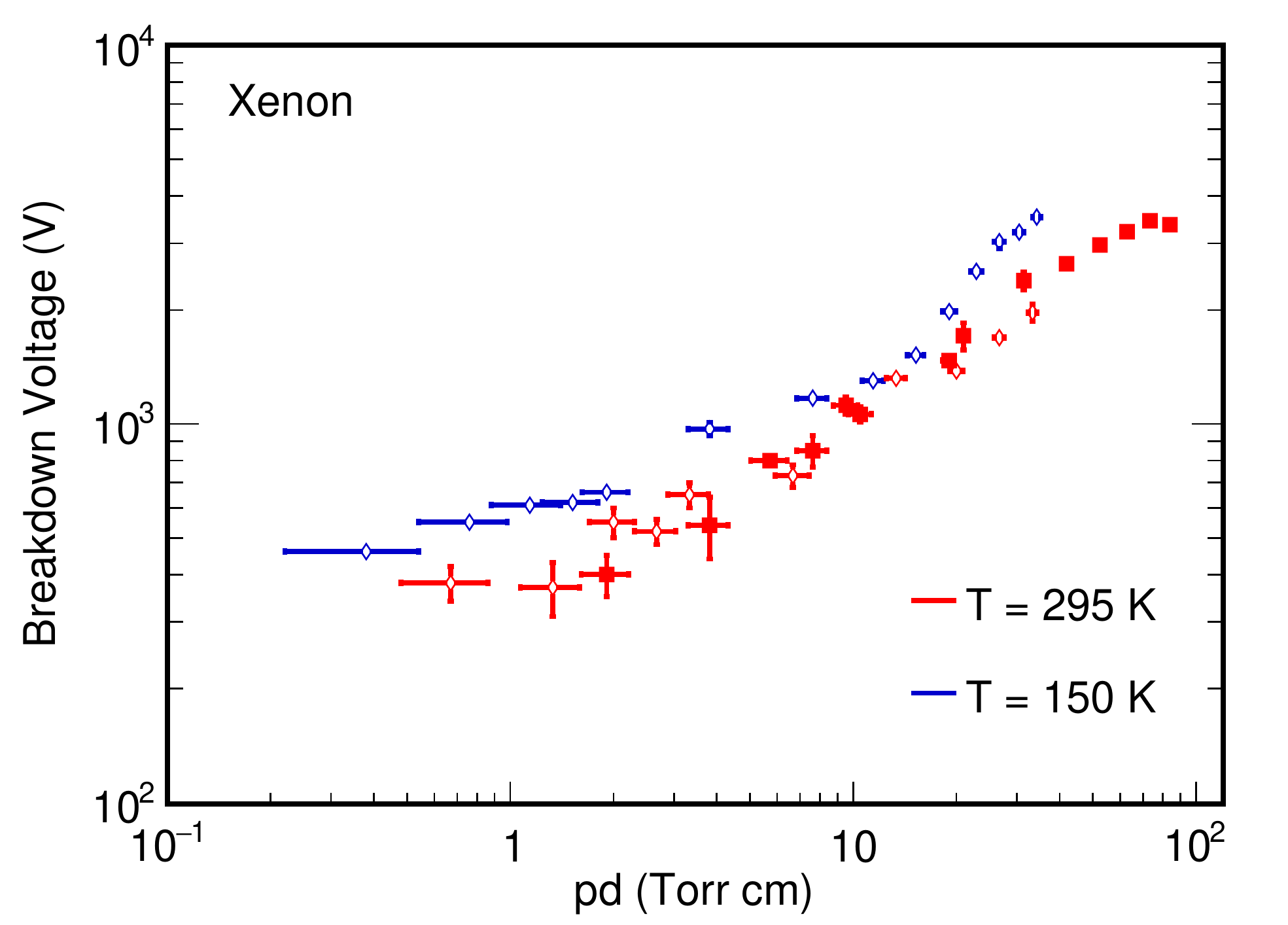}
 \includegraphics[width=0.9\columnwidth,,keepaspectratio=true]{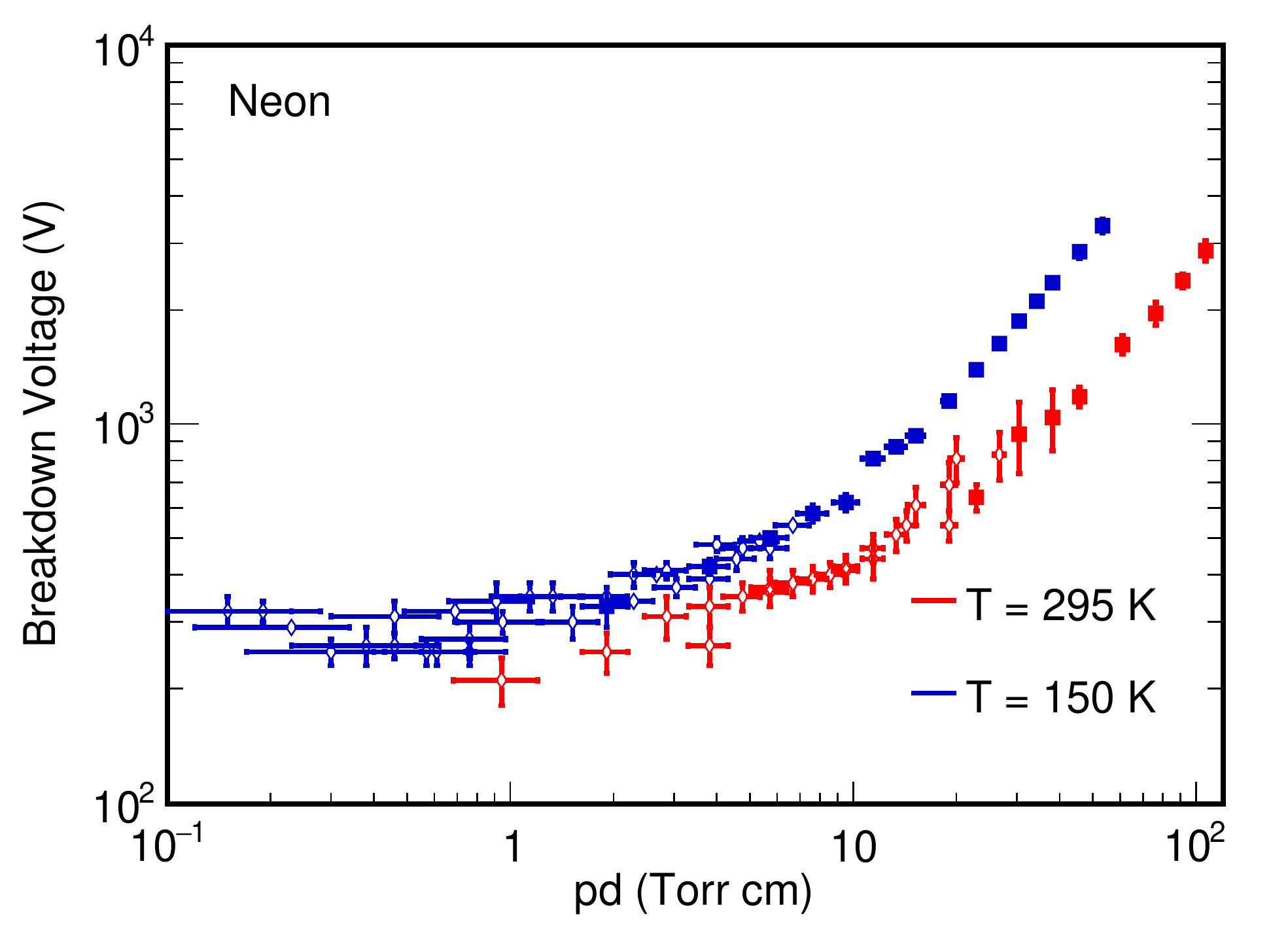}
 \caption{(color online)Break-through voltages for different gases and different gas temperatures. Different symbols
indicate
different pressure settings. The close squares indicate measurements around the normal pressure, open diamonds are
used for pressure settings from 70-300 Torr and triangles are used for the 7 Torr measurements (only in nitrogen). For
detailed pressure values see text. Red coloured data is measured at room temperature. Measurements at 150 Kelvin are
shown in blue and cyan stands for a gas temperature of 80 Kelvin (only nitrogen). }
 \label{Fig5}
\end{figure}

\begin{figure}[t!]

 \includegraphics[width=0.9\columnwidth,,keepaspectratio=true]{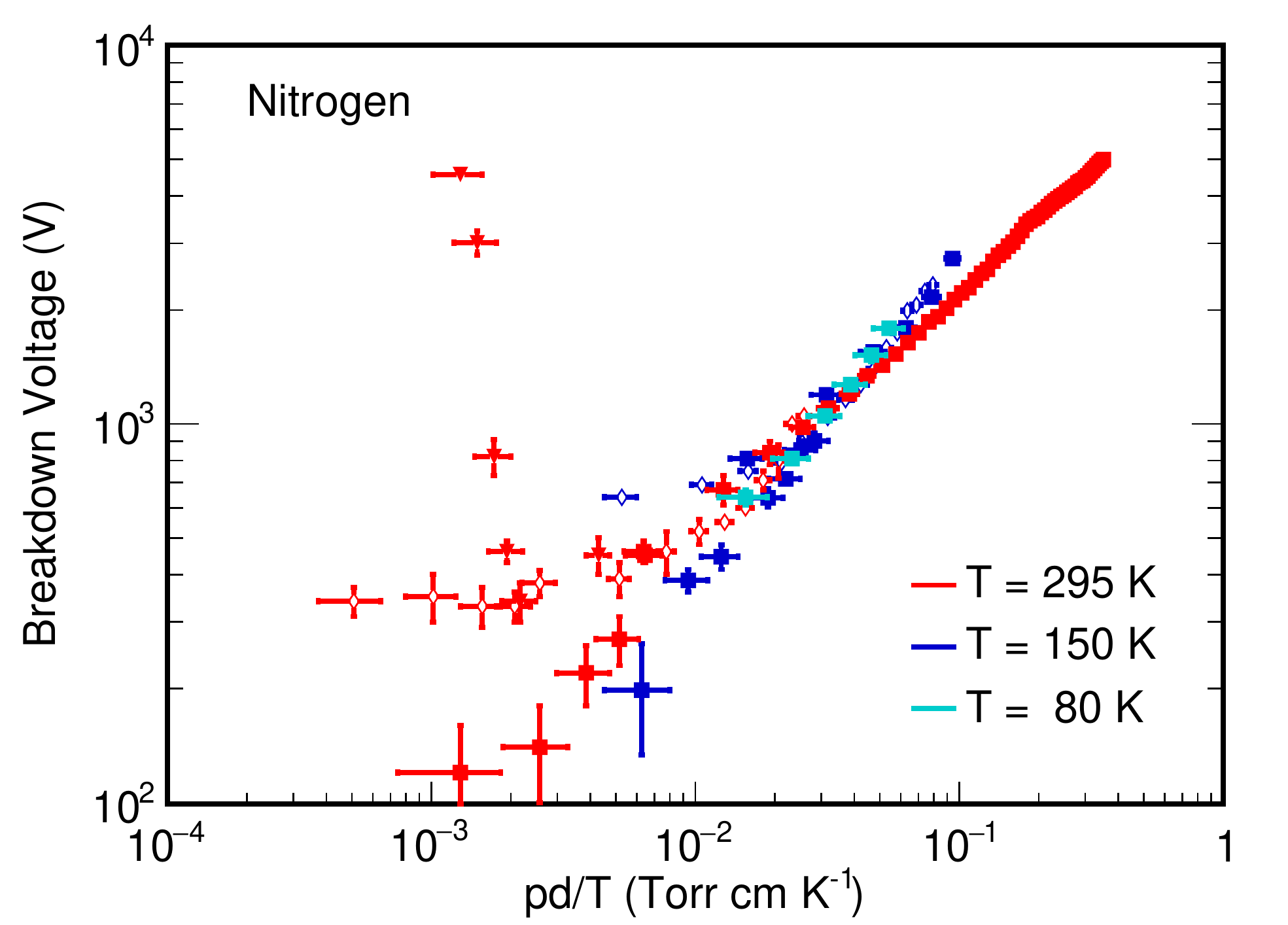}
 \includegraphics[width=0.9\columnwidth,,keepaspectratio=true]{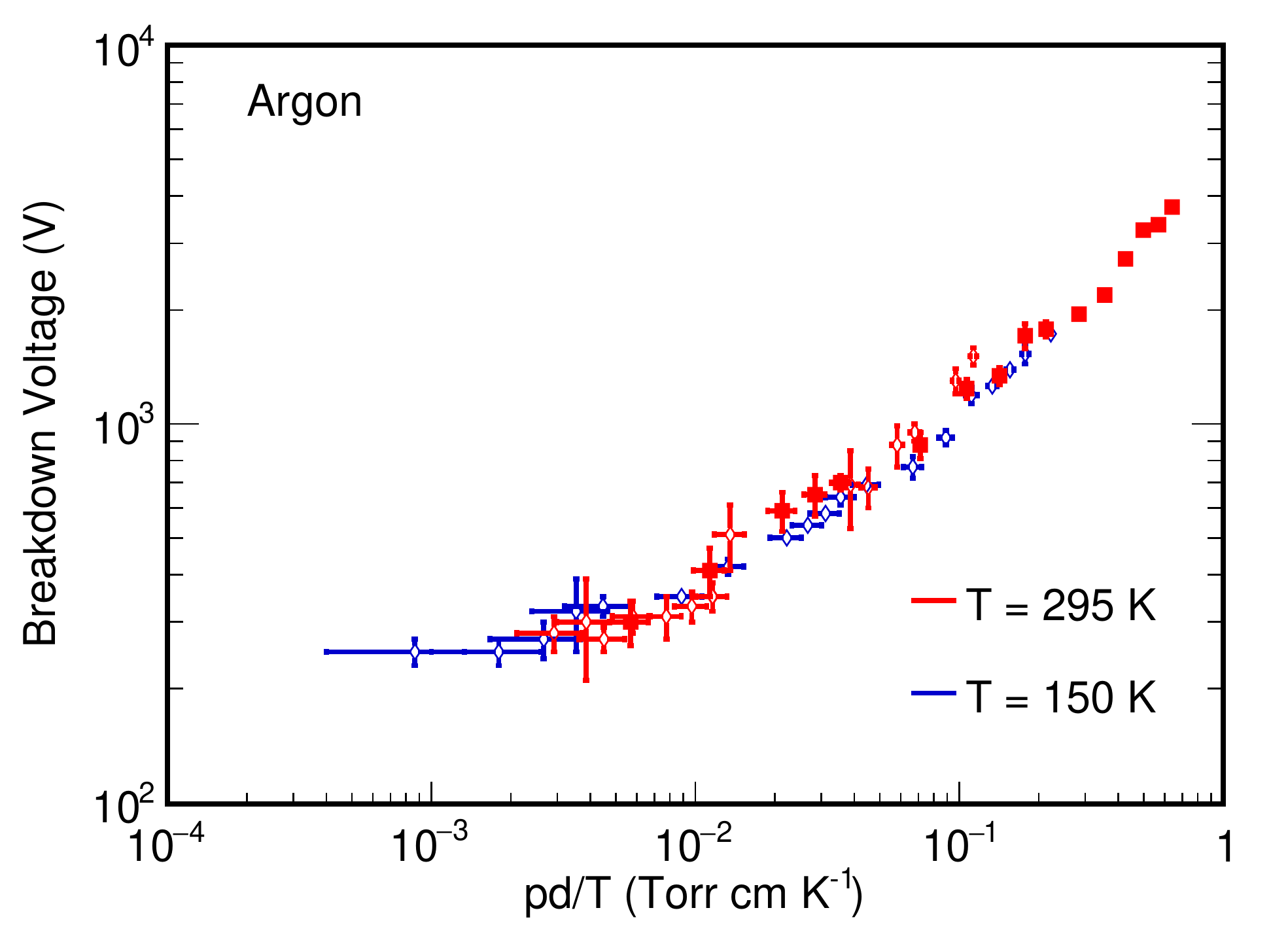}
 \includegraphics[width=0.9\columnwidth,,keepaspectratio=true]{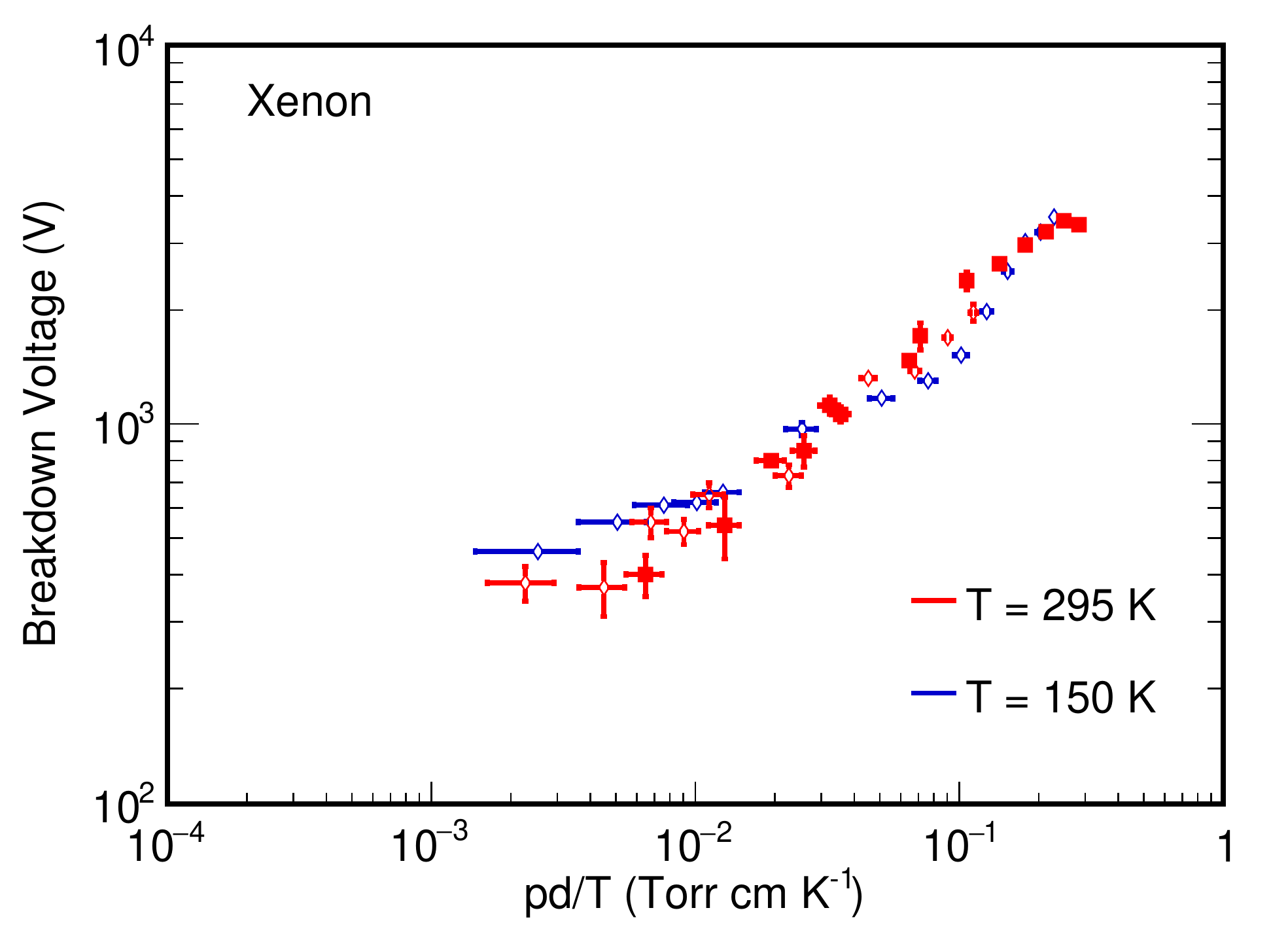}
 \includegraphics[width=0.9\columnwidth,,keepaspectratio=true]{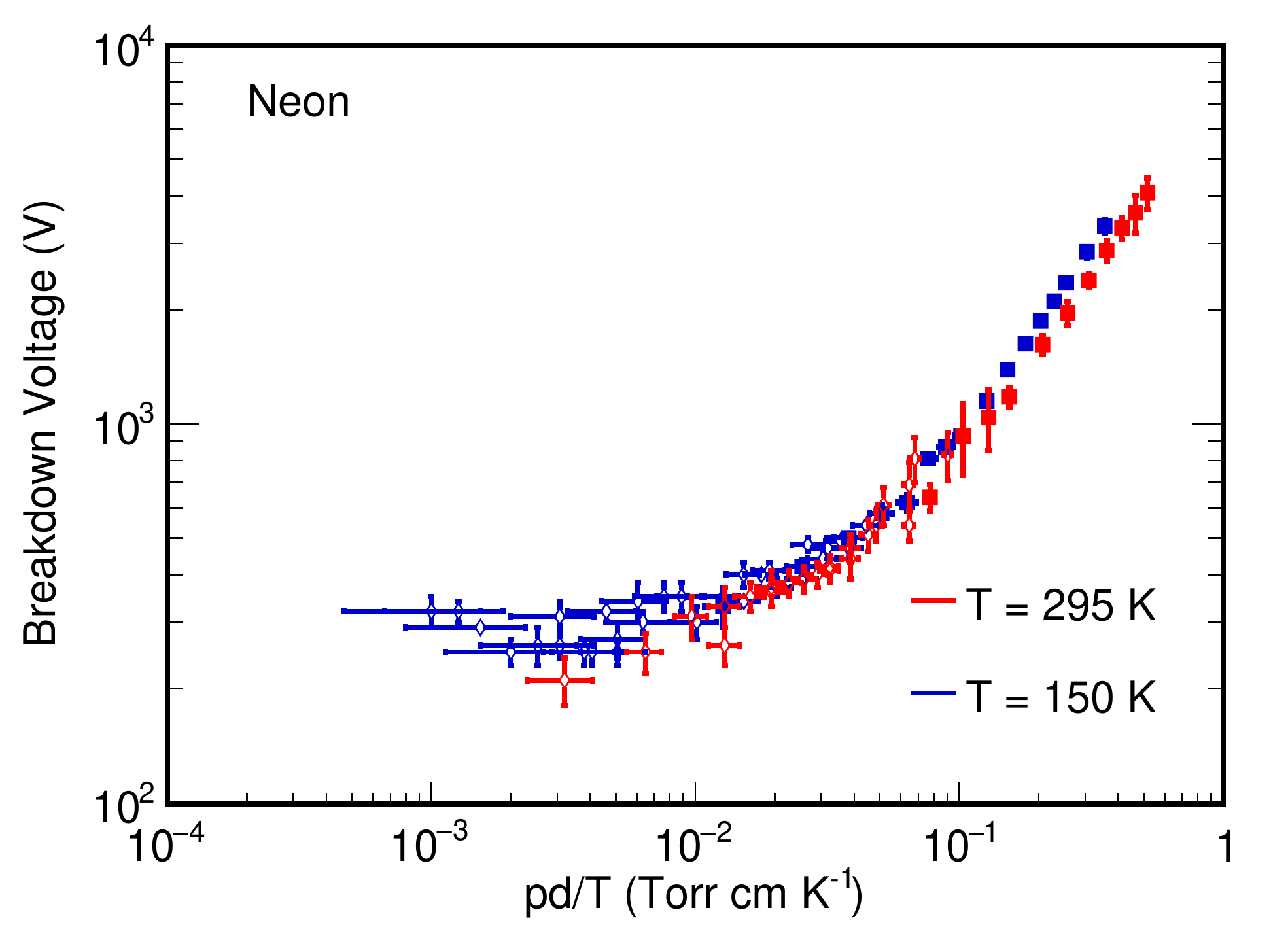}
 \caption{(color online)Break-through measurments plot in a generalized Paschen plot (see Sec.\,\ref{Sec_discussion})
based on
Eq.\,\ref{Eq_5}. The same color coding and symbols as in Fig.\,\ref{Fig5} are used. }
 \label{Fig6}
\end{figure}

Fig.\,\ref{Fig5} shows the data for the fill gases at different temperatures and pressures. The $V_B$ curves are
shifted
up for a given $pd$ when the temperature of the fill gas is lowered. In
Fig.\,\ref{Fig6}, $V_B$ is plotted vs. $pd/T$. This additional $T$ dependence is
motivated by the ideal gas law,
\begin{equation}
 p (A\cdot d) = n R T.
\label{Eq_3}
\end{equation}
Using $p$ and the volume, given here as area ($A$) times $d$, one can calculate the number of atoms $n$ at $T$
using the universal gas constant $R$. If $V_B$ for a pin is a function of the
mean free path of the electrons through the intervening gas, $V_B$ should be inversely proportional to the areal density
$n/A$. Transforming
Eq.\,\ref{Eq_3},
\begin{equation}
\frac{pd}{RT} = \frac{n}{A},
\label{Eq_4}
\end{equation}
the areal density is not only a function of $p$ and $d$ but also of $T$.
Neglecting the constant $R$, the factor on the left hand side of Eq.\,\ref{Eq_3}  is used in Fig.\,\ref{Fig6}. We
have found that the
general behavior of Paschen's law is still valid at low temperatures. Our
measurements also support the result of Ref.\,\cite{Gerhold1979} that Paschen's law should be generalized by introducing
a
temperature dependent factor arising from the ideal gas law.
As seen in Fig.\,\ref{Fig6} small deviations were observed. These deviations might arise because the gases are not
truly ideal. The better agreement in Neon, Xenon and Argon supports this hypothesis since we expect
these noble gases to be better described by the ideal gas law than Nitrogen.

\section{Summary}
In a series of measurements we have determined the breakthrough voltages for Nitrogen, Neon, Argon and Xenon.
In Nitrogen, deviations from Paschen's law were found when measuring at very small gaps. Indications
for the same behavior were found in the other gases. This is in agreement with existing literature.
For the first time, the general behavior of Paschen's law as a function of temperature has been systematically studied.
Therefore, the break-through voltages at various temperatures, distances and
pressures have been recorded. By using a classical plot of the Paschen curve, a shift in break-through voltage with
temperature was found and this shift can be explained well with a linear dependence of Paschen's law with temperature.
Therefore, we found that Eq.\,\ref{Eq_2} has to be used in a more general context,
\begin{equation}
 V_B = \frac{B\,\frac{pd}{T}}{ln(A'\,\frac{pd}{T})}.
\label{Eq_5}
\end{equation}
Both results are in agreement with previous findings of other groups. Existing fit parameters agree to the
results by applying Paschen's law to our data. Existing parameters $A$ and $B$ can still be used by applying a
temperature shift parameter $c = T\,/\,295K $. Eq.\,\ref{Eq_2} can be written then as:
\begin{equation}
 V_B = \frac{B_{T=295K}\,pdc}{ln(A'_{T=295K}\,pdc)}.
\label{Eq_5}
\end{equation}
The measurements with cold gases can be used to extrapolate our curves to lower temperature values. An additional
measurement in Nitrogen shows that such an extrapolation is reasonable.\\

The work presented in this manuscript has been performed on a open system where the gas can circulate or even is in
coexistence with a second phase. It has been shown that the behavior at room temperature is valid and shifts
can be explained based on the ideal gas law. The results of this work will influence designs of future high-purity
germanium detector systems and can help to improve existing designs.

\section{Acknowledgements}
We acknowledge the support of the U.S. Department of Energy through the LANL/LDRD Program.

\newpage

\bibliographystyle{apsrev}
\bibliography{paschen_paper}

\end{document}